# Non-equilibrium Tunneling Spectroscopy in Carbon Nanotubes

Yung-Fu Chen, Travis Dirks, Gassem Al-Zoubi*, Norman Birge*, Nadya Mason

Department of Physics and Materials Research Laboratory, University of Illinois at Urbana-Champaign, Urbana, IL 61801-2902, USA

*Department of Physics and Astronomy, Michigan State University, East Lansing, MI 48824-2320, USA

We report measurements of the non-equilibrium electron energy distribution in carbon nanotubes. Using tunneling spectroscopy via a superconducting probe, we study the shape of the local electron distribution functions, and hence energy relaxation rates, in nanotubes that have bias voltages applied between their ends. At low temperatures, electrons interact weakly in nanotubes of a few microns channel length, independent of end-to-end conductance values. Surprisingly, the energy relaxation rate can increase substantially when the temperature is raised to only 1.5 K.

Electronic transport in carbon nanotubes is relevant to a wide variety of applications, from nanoscale circuit elements [1] to quantum computers [2]. Nanotubes can be considered model one-dimensional systems [3] whose transport is strongly affected by electron-electron (e-e) interactions [4-8]. The influence of e-e interactions has often been observed via tunneling experiments, which can measure Luttinger liquid exponents [5,6], for example. However, these equilibrium experiments only probe interactions via the density of states (DOS) convolved with a Fermi distribution. In contrast, measurements of the *non-equilibrium* electron energy distribution function, $f(E)$, may directly determine e-e scattering and energy relaxation processes that are not apparent in the DOS [9-11]. In this Letter, we describe the first measurements of the non-equilibrium $f(E)$ in carbon nanotubes. Extracting the non-equilibrium $f(E)$ requires a means of biasing the electrons out of equilibrium while tunneling from a weakly coupled probe having a sharp feature in the DOS, such as a superconductor. Such a tunneling spectroscopy technique was first demonstrated on mesoscopic metal wires [12], where the non-equilibrium $f(E)$ and hence the scattering rate between quasiparticles were quantitatively determined [12]. Here, we perform similar experiments on carbon nanotubes, which, unlike the mesoscopic wires, are expected to be purely one-dimensional systems.

Our devices consist of metallic single-walled carbon nanotubes (diameters 1-3 nm, lengths 1.1-2.0 μm) having high conductance Cr/Au contacts at each end, a Pb superconducting tunneling probe in the middle, and a heavily doped Si substrate as a backgate (Figure 1(a)). The tunnel probes are separated from the nanotubes by a thin layer of $AlO_x$, and the tunneling resistances through the probes, $R_{tunnel}$ ~ 1-5 MOhm, are typically 10-100 times larger than the nanotubes' end-to-end resistances, $R_{end-to-end}$. Measurements were made through heavily filtered leads in a top-loading dilution refrigerator. Tunneling differential conductance measurements were performed by applying a sum of dc bias voltage $V$ and ac excitation voltage $V_{ac}$ to the superconducting probe, and a voltage $V_g$ to the back gate, while measuring the current $I$ at one of the nanotube end contacts (see Fig. 1(a)). For the non-equilibrium measurements, a non-zero DC voltage $U$ was applied across the nanotube end contacts. Note that because $R_{tunnel} \gg$ $R_{\text{end-to-end}}$, measurements of the tunneling differential conductance should not significantly perturb the electron distribution in the nanotube in comparison to $U$ [13]. For this manuscript, four devices, on separate chips, were measured in detail; all behaved similarly in non-equilibrium measurements.

The measurement regime is determined by the nanotube end-to-end conductance: Figures 2(a) and 2(b) show this for two different samples (A and B) as a function of gate voltage $V_g$, at temperature $T \approx 1.5$ K. Although $k_B T$ is smaller than the level spacing ($h v_F / L \sim 1$ meV) and charging energy ($e^2/2C \sim 2$ meV), the tube conductances do not pinch off to zero. In addition, although we see some oscillations as a function of $V_g$ and $U$, the conductance values reach $\sim e^2/h$ and any peaks are broadened. These observations imply that the coupling between the nanotube and the end contacts is strong, so that the measurements are taken in an open quantum dot regime. Thus, for the purpose of this experiment, we treat the nanotubes as having a continuous DOS that can be slightly modulated with $V_g$ and $U$ (see Fig. 1(b)).

When the nanotube is in equilibrium ($U = 0$), the tunneling current $I(V)$ through the superconductor/insulator/nanotube junction, in the open dot regime with bias $V$ across the junction, is given by

$$I(V) \approx \frac{1}{eR_T} \int_{-\infty}^{\infty} dE\, n_s(E+eV) n_{nt}(E)(f_{nt}(E) - f_s(E+eV)), \tag{1}$$

where $R_T$ is the tunnel resistance of the junction, $E$ is the energy relative to the Fermi energy of the nanotube, $n_s$ is the normalized BCS superconductor DOS, $n_{nt}$ is the normalized nanotube DOS, and $f_{nt}$ and $f_s$ are the Fermi distributions of the electrons in the nanotube and Pb probe, respectively. We extract the nanotube DOS $n_{nt}(E)$ from the equilibrium tunneling data by deconvolving Eq. (1). Although $n_{nt}(E)$ should have power law dependence as a function of $E$ if the nanotube is an ideal Luttinger liquid, this behavior is not usually seen in our samples (see Ref [14] and Supplementary info); it may be masked by the level discreteness, as the Thouless energy $\hbar v_F/L \sim 0.26$ mV is comparable to the measurement temperatures [15].

Figure 2(c) shows the differential tunneling conductance versus tunnel bias, $dI/dV$ vs $V$, of sample A at $V_g = 8.285$ V and $T = 1.3$ K. The expected [16] Pb superconducting gap $2\Delta \sim 2.6$ meV is evident as a zero conductance region centered around $V = 0$ between peaks at $V = \pm\Delta/e$. The peaks are BCS superconductor peaks convolved with the DOS of the nanotube and Fermi distributions of the Pb and the nanotube. The quality of the gap shows that our tunnel junction is relatively clean and non-invasive, and can indeed be used for energy-resolved spectroscopy. Above and below the gap region several more broadened peaks are also evident; these can be understood as tunneling peaks through multiple charge states in the open quantum dot (see Fig. 1(b)).

We next apply a non-zero voltage $U$ across the end contacts to drive the electrons in the nanotube out of equilibrium [12]: this introduces phase space for e-e scattering and allows us to measure the energy relaxation rates which may be due to this scattering. Because a complete theory for tunneling into a non-equilibrium one-dimensional system has not yet been formulated, we follow the precedent set in metals [12] and model our data using Eq. (1) with $f_{nt,U}(E)$ to be determined by experiment. In metal wires, $f_U(E)$ depends on the extent of electron energy relaxation in the wire, i.e. on the product of the inelastic scattering rate and the dwell time of an electron in the wire. This dependence can be understood by first considering two extreme cases: no inelastic scattering between electrons and strong inelastic scattering between electrons. In the first case, the non-interacting distribution function preserves the distributions of the two leads [17]: $f_0(E) = rf_L(E) + (1 - r)f_R(E)$, where $f_L(E) = (1 + \exp((E + eU)/k_BT))^{-1}$ and $f_R(E) = (1 + \exp(E/k_BT))^{-1}$ are the Fermi distributions in the left and right end contacts (with the right end grounded), respectively, and $r$ is the weight of $f_L(E)$ (determined by the tunneling rates into the two ends of the tube, the diffusivity of the nanotube, and the position of the superconducting probe [12,18]). When $eU \gg k_BT$, $f_0(E)$ is a two-step function. In the case of strong inelastic scattering between electrons a local electronic thermal equilibrium is created, with an effective temperature $T_{eff} \sim eU/k_B$ when $eU \gg k_BT$ [12,19]. This "hot" Fermi distribution is marked by a single broadened step. In general, the steady-state distribution function $f_{nt,U}(E)$ is between these two extreme cases and the shape of the distribution function reveals the extent of inelastic e-e scattering.

At $T = 1.3$ K, we see evidence of both strong and weak inelastic e-e scattering. Figure 3(a) shows $dI/dV(V)$ of sample A for various values of bias $U$ across the end contacts; Figures 3(b)-(c) show the same data taken at different gate voltages. The arrows in Fig. 3(a) indicate the superconducting peaks splitting at finite $U$. The decreasing height of the peaks at $V = \pm \Delta/e$ (compared with $U = 0$) and the newly developed peaks at $V = \pm \Delta/e + U$ are due to the fact that the states in the nanotube in the energy range of $(-eU, 0)$ are now partially occupied. From the differential form of Eq. (1) for finite U, we see that the clear separation between the peaks implies that the electron distribution now has two steps, at $E \sim 0$ and $-eU$, and thus that energy relaxation processes are weak. Figure 3(b), taken at a slightly different gate voltage, shows superconducting peaks that shift slightly, rather than separate, with $U$. In this case, instead of having a two-step-like electron distribution, $f_{nt,U}(E)$ has only one broad step in the energy range of $(-eU, 0)$, implying that energy relaxation processes are strong. Figure 3(c) shows behavior somewhere between 3(a) and 3(b). Although the back gate voltage tunes the nanotube conductance, we do not observe a clear correspondence between nanotube conductance and energy relaxation processes at finite $U$: data taken at $T \sim 1.5$ K with end-to-end conductance varied by up to a factor of 15 (near both peaks and valleys, for two samples at nine different gate voltages) shows little correlation between conductance values and the behavior of the superconducting peaks.

We do not see evidence of strong inelastic e-e scattering at temperatures well below 1.5 K. Figures 3(e)-(f) show typical $dI/dV(V)$ vs $U$ of sample B [20] at $T = 53$ mK: the superconducting peaks clearly split, implying two sharp steps in the electron energy distributions and hence weak scattering at finite $U$. This behavior can be compared to that at $T = 1.5$ K (Fig. 3(d)) where the lack of superconducting peak splitting implies a broad electron energy distribution and strong inelastic e-e scattering. Note that the superconducting peaks at 53 mK are much sharper than those at 1.3 K due to reduced thermal broadening of the distribution functions. The sharp splittings at low temperatures are independent of $V_g$ and tube end-conductance values: for example, Figs. 3(e) and 3(f) have similar peak splittings although the first is taken at a conductance peak and the second at a conductance valley (see insets). In general, we do not observe the effects of inelastic e-e scattering at these low temperatures, even in data taken at eight different gate voltage values where the tube conductance varies by a factor of 20.

In Fig. 4 we show the electron energy distribution functions extracted from the tunneling data in Fig. 3; the deconvolution was done using the differential form of Equation (1) (see Ref. [21] and Supplementary info). The shapes of the distributions are as expected from the behaviors of the peaks in Fig. 3 and the discussion above. The existence of double-step distribution functions for some of the curves (e.g., Fig. 4(a)) indicates that it is possible for the electrons to maintain their energy distribution across the lengths of the samples. However, surprisingly, the distribution functions are sometimes smeared and one-step-like near $T \sim 1.3$ K (c.f., Fig. 4(b)), even though $U \gg k_\text{B}T/e$. At lower temperatures, the distribution functions are always two-step like (Figs. 4(e)-(f)) and describe a system with weak energy relaxation. We note that the calculated distribution functions are very robust to small changes of $\Delta$ and $n_{\text{nt},U}(E)$ in the deconvolution process, implying that the shape of $f(E)$ is rather independent of the precise details of the fitting procedure. In addition, although some aspects of $f(E)$ are affected by the fitting procedure (e.g. a positive slope for $U = 0.5$ mV near $f(E) = 0.5$ in Fig. 4(e))—and even though the non-equilibrium form of Eq. (1) may not be exact for interacting systems—the overall shape of $f(E)$, double-step or rounded, is consistent with the qualitative behavior of the superconducting peaks in the raw $dI/dV(V)$ data.

Our data imply that inelastic scattering processes can be relatively weak in nanotubes. This may be because the typical electron dwell time in our tubes is short compared with that in a disordered metallic wire; $\tau = L/(v_\text{F}*t) \approx 50$ ps for a 2-micron long tube with nanotube Fermi velocity $v_\text{F} \approx 8 \times 10^5$ m/s and transmission $t = 0.05$ (corresponding to $R_\text{end-to-end} = 130$ k$\Omega$). However, even dwell times up to 400 ps ($R_\text{end-to-end} \sim 1$ M$\Omega$) do not lead to smearing of $f(E)$. The crossover from one-dimension to zero-dimensions may also limit inelastic scattering: in our samples the ballistic Thouless energy, $\hbar v_\text{F}/L \sim 0.26$ meV, is not much smaller than the typical bias voltage $U = 1.0$ mV. Our results may be consistent with theoretical predictions of no energy relaxation in out-of-equilibrium Luttinger liquid systems [11,22,23] unless the system is disordered [11]. Future experiments will examine the roles of tube length and disorder, and it is hoped that our results will also motivate further theoretical work. Overall, tunneling spectroscopy with a superconducting probe is a powerful new tool for characterizing e-e scattering in carbon nanotubes.

The authors wish to thank D. Esteve, L. Glazman, P.M. Goldbart, M. Kuroda, S. Lal, J. Leburton, L. Levitov, K. Matveev, W.-F. Tsai, and S. Vishveshwara for useful discussions. YFC, TD, and NM acknowledge support from the NSF under grant DMR-0605813 and the DOE Division of Materials Sciences under DE-FG02-07ER46453 through the Frederick Seitz Materials Research Laboratory. ALG and NB acknowledge support from the NSF under grants DMR-0405238 and 0705213. This research was carried out in part in the Center for Microanalysis of Materials, UIUC, which is partially supported by the DOE under DE-FG02-07ER46453 and DE-FG02-07ER46471.

References:

[1] P. Avouris, Z. H. Chen, and V. Perebeinos, Nature Nanotechnology **2**, 605 (2007).

[2] D. P. DiVincenzo, D. Bacon, J. Kempe, G. Burkard, and K. B. Whaley, Nature **408**, 339 (2000).

[3] C. Dekker, Physics Today **52**, 22 (1999).

[4] V. J. Emery, in *Highly Conducting One-Dimensional Solids*, edited by J. T. Devreese, R. P. Evrard and V. E. v. Doren (Plenum, New York, 1979), p. 247.

[5] M. Bockrath, D. H. Cobden, J. Lu, A. G. Rinzler, R. E. Smalley, T. Balents, and P. L. McEuen, Nature **397**, 598 (1999).

[6] Z. Yao, H. W. C. Postma, L. Balents, and C. Dekker, Nature **402**, 273 (1999).

[7] N. Y. Kim, P. Recher, W. D. Oliver, Y. Yamamoto, J. Kong, and H. Dai, Physical Review Letters **99**, 036802 (2007).

[8] F. Wu, P. Queipo, A. Nasibulin, T. Tsuneta, T. H. Wang, E. Kauppinen, and P. J. Hakonen, Physical Review Letters **99**, 156803 (2007).

[9] C. Lea and R. Gomer, Physical Review Letters **25**, 804 (1970).

[10] A. Komnik and A. O. Gogolin, Physical Review B **66**, 035407 (2002).

[11] D. A. Bagrets, I. V. Gornyi, and D. G. Polyakov, cond-mat, arXiv:0809.3166v1.

[12] H. Pothier, S. Gueron, N. O. Birge, D. Esteve, and M. H. Devoret, Physical Review Letters **79**, 3490 (1997).

[13] S. Gueron, (CEA-Saclay, 1997), Ph.D. thesis.

[14] We did not see these power laws in a sample where the tunneling probe was made normal by a magnetic field (see data in the supplementary information). In general, power law behaviors in single wall nanotube tunneling experiments at dilution refrigerator temperature have not yet been directly measured, as previous end-tunneling measurements were dominated by Coulomb blockade or Fabry-Perot effects.

[15] B. K. Nikolic and P. B. Allen, Journal of Physics-Condensed Matter **12**, 9629 (2000).

[16] I. Giaever, Physical Review Letters **5**, 147 (1960).

[17] K. E. Nagaev, Physics Letters A **169**, 103 (1992); Physical Review B **52**, 4740 (1995).

[18] H. Pothier, S. Gueron, N. O. Birge, D. Esteve, and M. H. Devoret, Zeitschrift Fur Physik B-Condensed Matter **104**, 178 (1997).

[19] V. I. Kozub and A. M. Rudin, Physical Review B **52**, 7853 (1995).

[20] Unfortunately, sample A was inadvertently destroyed before we were able to measure it at 50 mK.

[21] $n_{nt}(E)$ in the $U = 0$ case, and both $f_{nt,U}(E)$ and $n_{nt,U}(E)$ in the $U \neq 0$ cases were considered free parameters ($n_{nt,U}(E)$ is expected to change slightly with $U$ as quantized energy levels are shifted and the coupling to the leads also changes slightly in the open quantum regime); the fits to these functions minimize the least square difference between the measured and calculated values of $dI/dV(V)$. If $n_{nt,U}(E)$ is relatively flat we can fix $n_{nt,U}(E)$ to $n_{nt}(E)$ (the $U = 0$ case) and fit $dI/dV(V)$ reasonably well to get $f_{nt,U}(E)$; in this case $f_{nt,U}(E)$ has a similar shape

regardless of whether $n_{nt,U}(E)$ is fixed or a free parameter. If $n_{nt,U}(E)$ is not flat, we must let it vary for $U \neq 0$ in order to get a decent fit of $dI/dV(V)$; however, even in this case $n_{nt,U}(E)$ only seems to shift slightly with $U$, while maintaining its overall shape. In general, we find one unique solution for $n_{nt,U}(E)$ and $f_{nt,U}(E)$, independent of fitting conditions. See the supplementary information for more details.

[22] M. Khodas, M. Pustilnik, A. Kamenev, and L. I. Glazman, Physical Review B **76**, 155402 (2007).

[23] D. B. Gutman, Y. Gefen, and A. D. Mirlin, Physical Review Letters **101**, 126802 (2008).

Figure Captions:

Figure 1. (a) SEM image of a nanotube device and the measurement setup. Electrical measurements were performed by applying a sum of dc bias voltage $V$ and ac excitation voltage $V_{ac}$ to the superconducting probe, and voltage $V_g$ to the backgate, while measuring the current $I$ between the probe and one end of the nanotube (via a current preamplifier, which also acts as a virtual ground). A non-zero voltage $U$ could be applied to the other end contact. (b) Schematic diagram of an electron tunneling from a nanotube to a superconductor. The density of states (DOS) of the nanotube shows a modulation with single-particle energy spacing, as expected for an open quantum dot, while the superconducting DOS exhibits a BCS-like gap of $2\Delta$.

Figure 2. (a), (b) End-to-end differential conductance at $U = 1$ mV across samples A and B, respectively, as a function of gate voltage. (c) Tunneling differential conductance, $dI/dV$ of sample A from the superconductor into the nanotube, as a function of $V$ at $V_g = 8.285$ V. The blue arrow indicates the Pb superconducting gap size. Additional peaks at $V \sim -4.9, -2.8, 3.4$ mV are resonant tunneling peaks through the open quantum dot defined by the nanotube leads. (a) and (c) were taken at $T = 1.3$ K and (b) was at 1.5 K.

Figure 3. Tunneling differential conductance $dI/dV$ vs $V$ at multiple values of bias $U$ across the tube ends. (a) Sample A at $T = 1.3$ K, $V_{gate} = 8.660$ V. The peaks marked by black arrows are the superconducting peaks at $V = \pm\Delta/e$; the blue peaks marked by blue arrows are the superconducting peaks at $V = \pm\Delta/e + U$ (in this case $U = 1.5$ mV), (b)

Sample A at $T = 1.3$ K, $V_{gate} = 8.285$ V, (c) Sample A at $T = 1.3$ K, $V_{gate} = 8.070$ V, (d) Sample B at $T = 1.5$ K, $V_{gate} = -4.824$ V, (e) Sample B at $T = 53$ mK, $V_{gate} = -4.915$ V. The red peaks marked by red arrows are the superconducting peaks at $V = \pm\Delta/e$ and $V = \pm\Delta/e + U$ (in this case $U = 0.5$ mV), (f) Sample B at $T = 53$ mK, $V_{gate} = -4.460$ V. (a), (b) and (c) have the same legend, and (e) and (f) have the same legend.

Figure 4. Electron energy distributions calculated from the d$I$/d$V(V)$ data in Fig. 3. Two-step functions (a), (e), (f) imply limited e-e scattering, while broadened single-step functions (b), (d) imply strong e-e scattering. The dotted lines are non-interacting distribution functions $f_0(E)$ with $U = 1.0$ mV, $T = 1.3$ K, $r = 0.5$ in (a), $U = 1.0$ mV, $T = 1.3$ K, $r = 0.4$ in (c), and $U = 0.5$ mV, $T = 125$ mK, $r = 0.5$ in (f). Insets in (e) and (f): Zero-bias differential conductance across the nanotube as a function of $V_g$ (blue arrows indicate where d$I$/d$V(V)$ were taken).

Figure 1

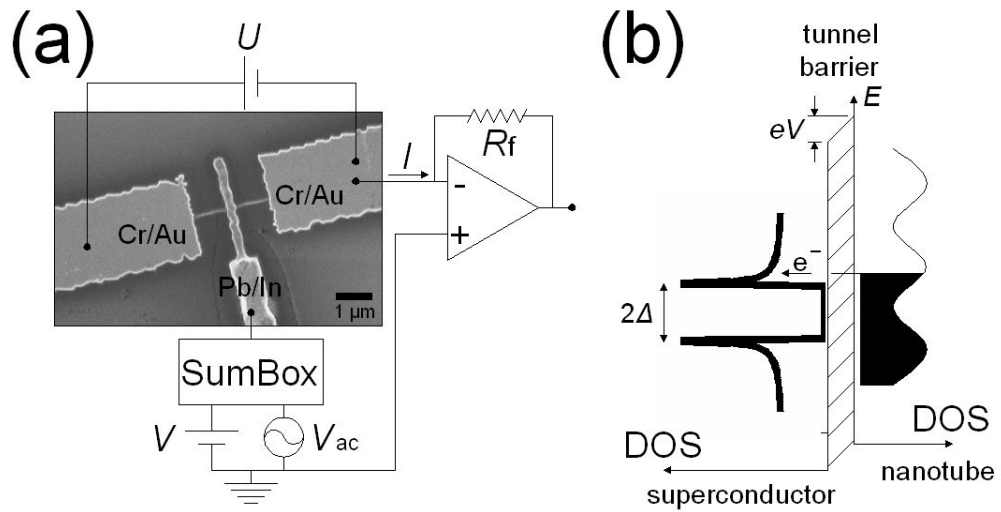

Figure 2

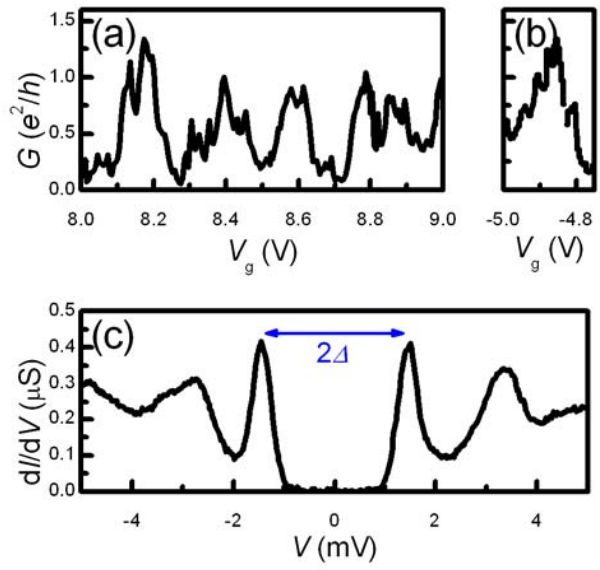

Figure 3

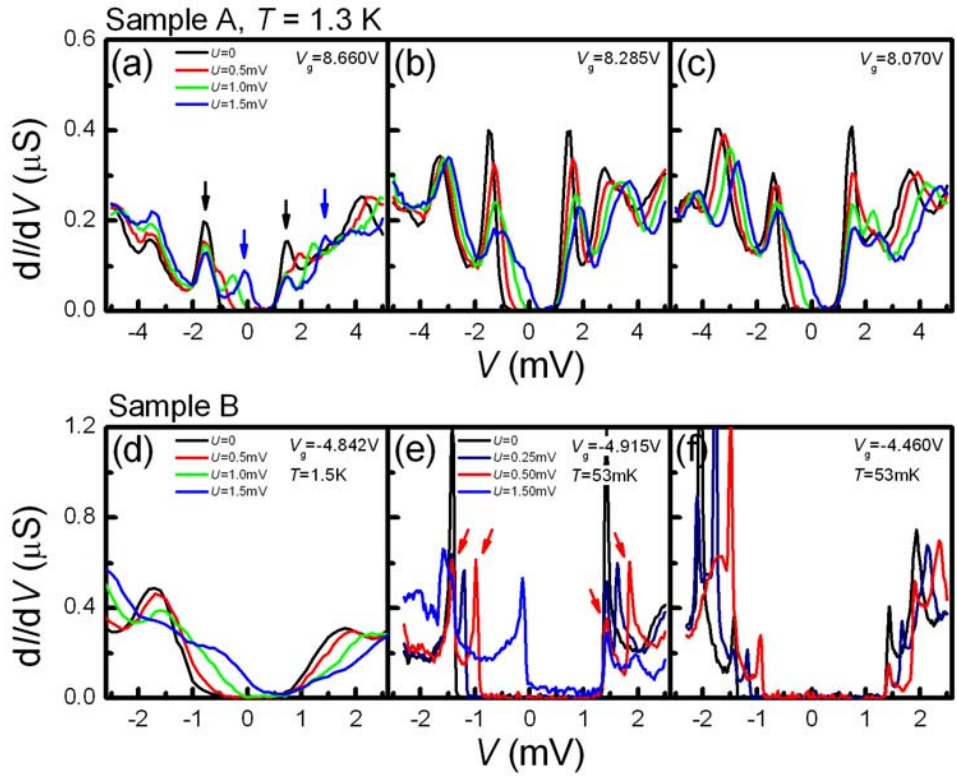

Figure 4

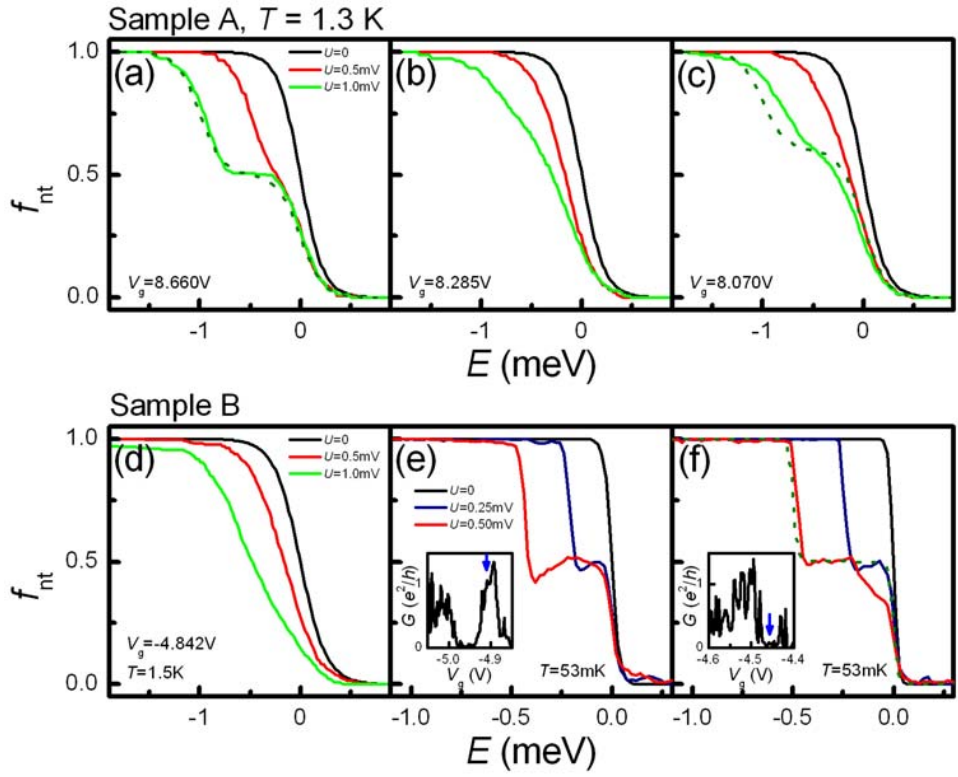